\documentclass{aa} 
\usepackage{graphicx}
\usepackage{txfonts}
\usepackage{multirow}
\usepackage{caption, threeparttable}
\usepackage{xcolor}
\usepackage{float}
\usepackage{amsmath}
\usepackage{subcaption}

\begin{document}

\newcommand{\teff}{$T_\mathrm{eff}$}
\newcommand{\logg}{$\log g$}
\newcommand{\feh}{[Fe/H]}
\newcommand{\microturb}{$\xi_\mathrm{micro}$}
\newcommand{\sife}{[Si/Fe]}
\newcommand{\mgfe}{[Mg/Fe]}
\newcommand{\afe}{[$\alpha$/Fe]}
\newcommand{\co}{CO($\nu =2-0$)\,}

\newcommand{\kssp}{{$\rm K_s$}\ }
\newcommand{\ks}{{$\rm K_s$}}
\newcommand{\ksnosp}{{$\rm K_s$}}
\newcommand{\jks}{{$\rm J-K_s$}\ }
\newcommand{\ejks}{{$\rm E(J-K_s)$} }

\newcommand{\mks}{$m_{K_{s}}$}

\newcommand{\water}{H$_2$O}
\newcommand{\invcm}{cm$^{-1}$}
\newcommand{\kms}{km\,s$^{-1}$}
\newcommand{\mic}{$\mu \mathrm m$}
\newcommand{\msun}{$\mathrm{M}_\odot$}

\defcitealias{Nandakumar:2023_I}{Paper\,I}
\defcitealias{Nandakumar:2023b_II}{Paper\,II}
\defcitealias{Nandakumar:24_III}{Paper\,III}
\titlerunning{NSD Abundance trends} 
\authorrunning{Ryde et al.}

   \title{Chemical Abundances in the Milky Way's Nuclear Stellar Disc }

   \author{N. Ryde
             \inst{1}
             \and
           G. Nandakumar
          \inst{2,1}
          \and
          R. Albarrac\'{i}n
          \inst{3,4}
        \and      
         M. Schultheis
           \inst{5}
         \and
         A. Rojas-Arriagada
         \inst{6,7,8}
         \and
         M. Zoccali
         \inst{7,9}
         }
   \institute{Division of Astrophysics, Department of Physics, Lund University, Box 118, SE-221 00 Lund, Sweden\\
              \email{nils.ryde@fysik.lu.se}
            \and
            Aryabhatta Research Institute of Observational Sciences, Manora Peak, Nainital 263002, India
            \and
   Max Planck Institute for Astronomy, D-69117 Heidelberg, Germany
   \and
   Fakultät für Physik und Astronomie, Universität Heidelberg, Im Neuenheimer Feld 226, 69120 Heidelberg, Germany
   \and
       Université Côte d’Azur, Observatoire de la Côte d’Azur, Laboratoire Lagrange, CNRS, Blvd de l’Observatoire, 06304 Nice, France
   \and
       Departamento de F\'isica, Universidad de Santiago de Chile, Av. Victor Jara 3659, Santiago, Chile
   \and
Millennium Institute of Astrophysics, Avenida Vicuña Mackenna 4860, Macul, Santiago 82-0436, Chile
   \and
Center for Interdisciplinary Research in Astrophysics and Space Exploration (CIRAS), Universidad de Santiago de Chile, Santiago, Chile
\and
Instituto de Astrofísica, Pontificia Universidad Católica de Chile, Av. Vicuña Mackenna 4860, 782-0436 Macul, Santiago, Chile}
       
   \date{Received ; accepted }

 
  \abstract
   {
    The Nuclear Stellar Disc (NSD) is a rotating, disc-like structure in the Galactic Center, believed to have a distinct star formation history and a predominantly old stellar population. However, its formation history and evolutionary links to other structures in the Galactic Center remain uncertain. Studying the chemical evolution of the NSD could provide new insights into this region and key epochs in the Milky Way evolution, yet such studies remain rare.}
{This study aims to present the first comprehensive chemical census of the NSD by deriving abundance trends for 18 elements in nine M giants  in the metallicity range  of $-1.0<$\feh$<+0.5$. By comparing these trends with those of other Galactic populations -- including the Nuclear Star Cluster (NSC), the inner bulge, as well as the thin and thick discs -- we seek to understand the chemical relationships between these structures.}  
{To mitigate the extreme optical extinction along the line of sight, we obtained high-resolution H- and $\mathrm{K}_{s}$-band spectra of NSD stars using the IGRINS spectrometer mounted on the Gemini South telescope. The observed M giants were analyzed consistently with stars from the comparison populations to minimize systematic uncertainties.}
{The abundance trends of NSD stars exhibit strong similarities with those of the inner-bulge and NSC populations across a broad range of elements with different chemical evolution histories. The trends for $\alpha$-elements, Al, Cr, Mn, Co, Ni, Cu, Zn, and neutron-capture elements align closely with the local thick-disc behavior at subsolar metallicities. At super-solar metallicities, most elements follow the NSC and inner-bulge trends. Sodium is the only element exhibiting a distinct trend, with enhanced abundances in the NSD and NSC compared to both thin-disc and inner-bulge stars.}
{The chemical similarity of most of the 18 elements investigated, including Na, suggests that the NSD likely shares an evolutionary history with the NSC and possibly the inner-disc sequence. Further studies are required to determine potential evolutionary links to the complex stellar system Liller 1 and metal-rich globular clusters. We find no evidence of typical globular cluster abundance signatures in our NSD stars with subsolar metallicities. Our study demonstrates the feasibility of obtaining high-quality abundance data even in highly dust-obscured regions of the Milky Way, paving the way for future surveys.}


   \maketitle
%

\section{Introduction}
\label{sec:intro}
\vspace{-5pt}

The Galactic Center is a unique environment located near the Milky Way's  supermassive black hole,  characterized by high gas densities and turbulence in the Central Molecular Zone (CMZ), along with ongoing star formation. The Nuclear Stellar Disc (NSD) and the Nuclear Star Cluster (NSC) are the dominant stellar structures near the Galactic Center, and are thought to host distinct stellar populations that are predominantly old, metal-rich, and shaped by unique star formation histories \citep{schodel:20,lara:20,hvs}. A priori, there is no obvious reason to expect that the stellar populations in the Galactic Center would share significant similarities with those in the Milky Way’s bulge.

The NSD is a flat, rotating, and kinematically cold disc-like structure (see, e.g., Schönrich et al. 2015\nocite{schonrich:15}; for an alternative interpretation, see  Zoccali et al. 2024\nocite{zoccali:24}) that surrounds the much smaller NSC. The NSD has a break radius of $\sim90$\,pc, a vertical height of $\sim45$\,pc, and a total stellar mass of $1\times 10^9\,\mathrm M_\odot$ \citep[see, e.g.,][]{launhardt:02,bland:16}. The NSD resides within the CMZ, which spans  $\sim250\times50$\,pc$^2$ \citep{henshaw:23,jwst_nsc_white:23} and is embedded in the Galactic bulge/bar \citep{sormani:22}. The CMZ is the largest reservoir of dense gas in the Milky Way \citep{baba:20,sormani:22}, with significant gas inflows channeled by the Galactic bar \citep[see, e.g.,][]{gadotti:20}.

NSDs are common in spiral galaxies and likely form following the dynamic settling of the Galactic disc and the buckling of the bar \citep{baba:20}, i.e., after the bar has formed \citep[see, e.g.,][]{desafreitas:22}. The Milky Way’s NSD is thought to have formed through inside-out star formation occurring in nuclear rings \citep{sormani:22}. Although the bulk of the NSD’s stellar population is believed to have formed more than 8 Gyr ago \citep{matsunaga:11,lara:20}, it appears to have experienced a significant starburst approximately 1 Gyr ago \citep{lara:20}.

The star formation histories (SFHs) in this region have varied over time, are not quasi-continuous, and appear distinct in the NSD and NSC \citep{schodel:20,lara:21}. Despite differences in stellar populations and formation histories, \citet{lara:23} suggest -- based on kinematics and metallicity gradients -- that there is a smooth transition between the NSD and NSC, potentially indicating that they are parts of the same structure. Additionally, \citet{sanders:24} propose an evolutionary link between the NSD and NSC, inferring that both systems are predominantly old, though generally younger than the main bar/bulge. The metal-poor population of the NSD may have originated from clusters in the inner Galaxy or could consist of interlopers from the Galactic bar.


Research on the Galactic Center, including the NSC and the surrounding NSD, is an active field of study, with for example galaxy modelers increasingly attempting to simulate their formation using high-resolution zoom-in simulations \cite[see, e.g.,][]{zoom2,zoom3}. Due to the extreme observational challenges (with up to 30 magnitudes of optical extinction), research on the Galactic Center has traditionally focused on theoretical modeling, complemented by observations primarily using photometry and low-resolution spectroscopy to study the dynamics of the systems and metallicity distributions \citep[see, e.g.,][]{fritz:21,feld:25}. While a few isolated abundance measurements have been made, no comprehensive studies of its chemical evolution have been conducted until recently -- in contrast to the well-studied discs and bulge populations of the Milky Way. Such chemical evolution studies would add a new dimension to our understanding of the Galactic Center.


Thus, due to the significant extinction toward the Galactic Center, abundance studies of the NSD remain scarce. Determining abundance trends has been extremely challenging, requiring large telescopes and high-resolution observations of M giants, which are particularly difficult to analyze. Nonetheless, \citet{ryde:15,ryde:2016_metalpoor,ryde:2016_bp2,Nandakumar:18,Nieuwmunster:2023} presented [Mg/Fe] and [Ca/Fe] trends as a function of metallicity for nine giants belonging to the old NSD population, located $2.5-5.5$\,\arcmin\ north of the Galactic Center (corresponding to a projected galactocentric distance of $5-10$\,pc). Their findings indicated thick-disc-like abundance patterns for metal-rich stars in this region. Similar results were reported by \citet{Schultheis:2020} for a handful of NSD stars observed in the APOGEE survey \citep[Apache Point Observatory Galactic Evolution Experiment;][]{Holtzman:2018,Jonsson:2018}. In the first determination of [Si/Fe] for old stars in the NSD, \citet{thorsbro:2020} found a disc-like trend at subsolar metallicities, whereas stars at supersolar metallicities exhibited enhanced abundances.


Since different groups of elements trace distinct nucleosynthetic pathways, each with its own evolutionary timescale, abundance trends as a function of metallicity for as many elements as possible, covering a wide range of nucleosynthetic processes, 
offer valuable insights into the characteristics of a stellar population \citep[see, e.g.,][]{matteucci:12,Matteucci:2021,manea:23}, such as that of the NSD. These trends can differ across stellar populations with varied evolutionary histories. When combined with Galactic chemical evolution models, they help constrain key processes, including star-formation history, nucleosynthetic pathways, and the role of gas infall.


Today, studying chemical evolution is feasible even in the most dust-obscured regions of the Galaxy, such as the Galactic Center, thanks to high-resolution infrared spectrographs like the Immersion GRating INfrared Spectrograph \citep[IGRINS;][]{Yuk:2010, Wang:2010, Gully:2012, Moon:2012, Park:2014, Jeong:2014}. This advancement enables the investigation of chemical relationships across  Galactic populations, including those in the NSC, NSD, bulge, thick disc, and thin disc of the Milky Way. For instance, \citet{ryde:25} and \citet{NSC_all:25} present the first chemical census of 19 elements for the old, metal-rich stellar population in the NSC, demonstrating its similarities to inner-bulge populations.

In this paper, we present an unprecedented dataset of elemental trends for 18 elements successfully derived for stars in the NSD. To investigate the relationships between different Galactic populations and to identify differences that may reflect distinct chemical evolution, minimizing systematic uncertainties is essential. Therefore, we compare our abundance trends for NSD stars with those of the solar neighborhood \citep{Nandakumar:2023,Nandakumar:24_21elements}, as well as with those of the inner-bulge population investigated by \citet{nandakumar:24}, located $1^\circ$ north of the Galactic Center, and those of the NSC \citep{ryde:25,NSC_all:25}. All these samples were observed using the same instrumental setup and analyzed with a consistent methodology, including an identical stellar parameter scale and mostly the same spectral lines. This uniform approach enables direct, element-by-element, and line-by-line differential comparisons, ensuring a homogeneous analysis. Systematic uncertainties are expected to impact all four samples similarly. For all these populations, we derived abundance trends as a function of metallicity for a range of elements, including $\alpha$-elements, odd-Z elements, iron-peak elements, and neutron-capture elements. Incorporating such a broad range of elements is crucial for achieving a comprehensive chemical perspective on the different components. We are now well-positioned to analyze the chemistry of the NSD in a comparative study alongside the Galactic discs, inner bulge, and NSC.

 \begin{figure*}
  \includegraphics[trim=3.5cm 0cm 4cm 0cm,clip,width=0.45\textwidth, angle=-90]{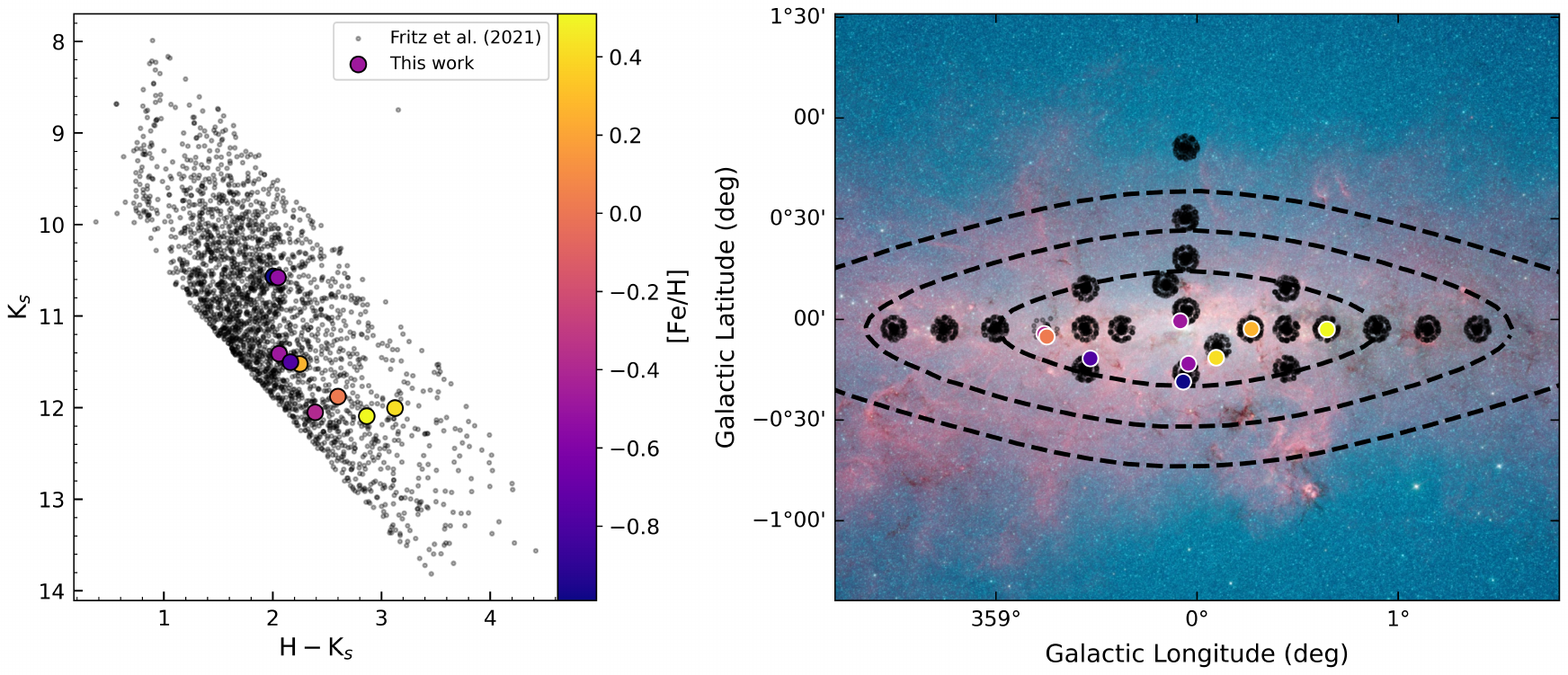}
  \caption{\textit{Left}: K$_s$ vs (H - K$_s$) diagram of our targets, colored by metallicity, superimposed on stars from the \citet{fritz:21} KMOS catalog. \textit{Right}: Distribution of our targets in the sky as colored circles, again colored by metallicity. Black dashed lines show the density contours of the NSD model from \citet{sormani:22} over an image from the GLIMPSE survey \citep{GLIMPSE,GLIMPSE2}. Black dots represent the sample from \citet{fritz:21}.}
  \label{fig:map}%
\end{figure*}

\begin{table*}
\caption{Observational data of the observed M giants in the NSD.} \label{table:obs}
\centering
\begin{tabular}{l  c c c c c c c}
\hline
\hline
Star Name & R.A.  & DEC & Obs. date & H$_\mathrm{2MASS}$  & K$_\mathrm{2MASS}$      & S/N$_\mathrm{H}$ & S/N$_\mathrm{K}$ \\
&   (hh:mm:ss) & (dd:mm:ss)  & UT & (mag) & (mag) &  \multicolumn{2}{c}{per resolution element}  \\ 
\hline
 NSD\_1  & 17:45:27.45 & -29:00:44.14 & 2024-04-09 &13.47  & 11.41  &  40  &  160	 \\ 
 NSD\_8  & 17:46:35.48 & -28:57:15.59 & 2024-04-03 & 15.13 & 12.00  &  40  &  150	 \\ 
 NSD\_12 & 17:46:27.02 & -28:43:48.32 & 2024-04-09 &  13.77 & 11.52 &  30  &  150	 \\ 
 NSD\_16  & 17:46:40.14 & -29:09:22.28 & 2024-04-02 & 12.57 & 10.57  &  60  &  210	 \\ 
 NSD\_19  & 17:46:22.83 & -29:05:15.07 & 2024-04-03 & 12.63 & 10.58 &  40  &  130	 \\ 
 NSD\_25  & 17:47:20.89 & -28:24:37.40 & 2024-04-02 &  14.96 & 12.09 &  20  &  140	 \\ 
 NSD\_30  & 17:44:04.98 & -29:37:17.33 & 2024-04-01 & 14.44 &12.05  &  30  &  130	 \\ 
 NSD\_31  & 17:44:10.03 & -29:37:04.80 & 2024-04-01 & 14.48 & 11.88    &  30  &  130	   \\ 
 NSD\_33 & 17:45:06.83 & -29:29:24.04 & 2024-04-03 & 13.66 & 11.50  &  40  &  120	 \\ 
\hline    

\end{tabular}
\end{table*}

\begin{table*}
\caption{Stellar Parameters for the investigated NSD Stars}\label{table:parameters}
\centering
\begin{tabular}{lcccccccc}
\hline\hline
Star & $T_\mathrm{eff}$ & $\log g$ & [Fe/H] & $\xi_\mathrm{micro}$ & [C/Fe] & [N/Fe] & [O/Fe] \\
 & (K) & (cgs) & (dex) & (km\,s$^{-1}$) & (dex) & (dex) & (dex) \\
\hline
NSD\_1 & 3840 & 1.0 & -0.48 & 2.0 & -0.45 & 1.37 & 0.36 \\
NSD\_8 & 3770 & 1.5 & +0.42 & 1.5 & -0.02 & 0.31 & -0.04 \\
NSD\_12 & 3640 & 1.1 & +0.26 & 1.6 & -0.00 & 0.33 & 0.03 \\
NSD\_16 & 3620 & 0.2 & -0.99 & 2.5 & 0.14 & 0.48 & 0.60 \\
NSD\_19 & 3580 & 0.5 & -0.53 & 2.0 & -0.38 & 1.35 & 0.42 \\
NSD\_25 & 3460 & 1.0 & +0.51 & 1.5 & -0.13 & 0.60 & -0.07 \\
NSD\_30 & 3510 & 0.5 & -0.39 & 2.0 & 0.19 & 0.21 & 0.33 \\
NSD\_31 & 3470 & 0.6 & +0.02 & 2.3 & 0.14 & 0.27 & 0.15 \\
NSD\_33 & 3670 & 0.5 & -0.79 & 2.4 & 0.21 & 0.25 & 0.51 \\
\hline
\end{tabular}
\end{table*}

\begin{table*}
\caption{Elemental abundances for the investigated NSD stars}
\label{tab:element_abundances}
\centering
\begin{tabular}{lccccccccc|c}
\hline\hline
Element & NSD\_1 & NSD\_8 & NSD\_12 & NSD\_16 & NSD\_19 & NSD\_25 & NSD\_30 & NSD\_31 & NSD\_33 & Uncertainty$^{(a)}$ \\

\text{[Fe/H]} & -0.48 & 0.42 & 0.26 & -0.99 & -0.53 & 0.53 & -0.39 & 0.02 & -0.79 & $\pm 0.10$\\
\hline  
\text{[F/Fe]} & $*^{(b)}$ & 0.25 & 0.28 & -0.37 &  $*^{(b)}$ & 0.15 & -0.07 & 0.21 & -0.20 & $\pm 0.06$ \\
\text{[Na/Fe]} & 0.63 & 0.64 & 0.54 & 0.02 & 0.36 & 0.51 & 0.19 & 0.54 & 0.27 & $\pm 0.05$\\
\text{[Mg/Fe]} & 0.14 & -0.08 & -0.09 & 0.29 & 0.02 & -0.12 & 0.19 & -0.09 & 0.34 & $\pm 0.09$\\
\text{[Al/Fe]} & 0.29 & -0.02 & -0.07 & 0.02 & 0.10 & 0.09 & 0.14 & -0.06 & 0.10  &$\pm 0.05$\\
\text{[Si/Fe]} & 0.15 & -0.06 & 0.07 & 0.43 & 0.18 & 0.02 & 0.18 & 0.05 & 0.32  &$\pm 0.10$\\
\text{[S /Fe]} & 0.31 & -0.19 & -0.31 & 0.17 & 0.25 & -0.55 & 0.04 & -0.19 & 0.49 & $\pm 0.12$\\
\text{[K /Fe]} & - & - & - & - & - & - & - & - & -  & - \\
\text{[Ca/Fe]} & 0.24 & -0.08 & -0.03 & 0.31 & 0.15 & -0.15 & 0.11 & -0.02 & 0.19 & $\pm 0.07$\\
\text{[Ti/Fe]} & 0.46 & 0.16 & 0.14 & 0.53 & 0.42 & 0.15 & 0.27 & 0.09 & 0.46 & $\pm 0.10$\\
\text{[Cr/Fe]} & -0.29 & -0.11 & 0.01 & -0.13 & - & - & - & - & -  & $\pm 0.05$\\
\text{[Mn/Fe]} & -0.08 & 0.19 & 0.24 & -0.17 & -0.01 & - & 0.05 & 0.36 & 0.01 & $\pm 0.11$ \\
\text{[Co/Fe]} & 0.14 & 0.03 & -0.08 & 0.08 & -0.01 & -0.23 & -0.03 & 0.07 & 0.16 & $\pm 0.07$\\
\text{[Ni/Fe]} & 0.04 & 0.15 & 0.12 & 0.18 & 0.01 & 0.17 & 0.09 & 0.10 & 0.11 & $\pm 0.07$ \\
\text{[Cu/Fe]} & 0.07 & 0.21 & 0.00 & -0.11 & 0.02 & 0.40 & 0.00 & - & -  & $\pm 0.08$\\
\text{[Zn/Fe]} & 0.47 & 0.16 & 0.30 & - & - & - & - & 0.15 & 0.62  &$\pm 0.06$\\
\text{[Ba/Fe]} & - & -0.09 & -0.15 & 0.38 & 0.27 & 0.00 & 0.02 & - & 0.75  & $\pm 0.15$\\
\text{[Ce/Fe]} & 0.19 & -0.13 & -0.11 & 0.16 & 0.25 & -0.35 & -0.20 & -0.12 & 0.18  &$\pm 0.08$\\
\text{[Nd/Fe]} & 0.62 & -0.02 & -0.24 & 0.31 & - & - & 0.28 & 0.18 & 0.48  & $\pm 0.06$\\
\text{[Yb/Fe]} & 0.42 & - & 0.42 & 0.76 & 0.37 & 0.35 & 0.54 & 0.29 & 0.14  & $\pm 0.09$ \\
\hline
\end{tabular}
\tablefoot{
\tablefoottext{a}{Typical uncertainties due to uncertainties in stellar parameters for a typical star.}\\
\tablefoottext{b}{These lines are very weak, making the fluorine abundances unmeasurable.}}

\end{table*}









\section{Observations and data reduction}
\label{sec:obs}

We obtained high-resolution spectra of nine M giants in the NSD using the IGRINS spectrograph mounted on the Gemini South telescope \citep{Mace:2018}. The IGRINS spectra, with a resolution of $R\sim45,000$, span the entire H and K bands ($1.45-2.5$\,\mic). The observations were conducted  between April 1 and April 8, 2024, under program GS-2024A-Q-216 (PI: Albarrac\'{i}n. 

Our nine target stars were selected from the KMOS study of the NSD by \citet{fritz:21} and are shown in Fig.~\ref{fig:map}.
In the left panel of the figure, we display our targets in the color-magnitude diagram against the background of the full KMOS sample. In the right panel,  our targets are plotted on the NSD model by \citet{sormani:22}. The contours in the figure illustrate the ratio between the surface density (on the plane of the sky) of the NSD model and a model of the Galactic bar, indicating the fraction of stars belonging to the NSD at each position in the sky. Our targets are situated in the inner regions, where NSD stars account for at least $75\%$ of the stellar population,  minimizing potential contamination from the bulge. The membership to the NSD is based on proper motions and detailed orbital calculations; 
We used proper motions derived from the VIRAC2 \citep{Smith2018} photometric and astrometric reduction of the VVV data \citep{Minniti2010} with small uncertainties, i.e. $\rm \mu_{l,err} < 0.6\,mas/yr$, and $\rm \mu_{b,err} < 0.6\,mas/yr$. Combined with radial velocities, orbital parameters were determined using the software package AGAMA \citep{Vasiliev2019}, assuming the most realistic gravitational potential for the NSD: a combined  potential of the NSC, the NSD, and a rotating bar \citep[][]{Nieuwmunster2024}. We assumed a distance of 8.2\,kpc for the stars but relaxed this assumption by using MCMC simulations within a spread of 100\,pc. All derived orbits show apocentric radii and $z_\mathrm{max}$ values typically for the NSD, making us confident that our stars are indeed members to the NSD.

The coordinates of our stars are provided in Table \ref{table:obs}, along with their H and K magnitudes \citep[from][]{Nishiyama2013}, as well as the signal-to-noise ratios per resolution element (S/N). The K magnitudes range from 10.5 to 12.0.  
The H band S/N are generally a factor of 3 lower compared to the K band.

The IGRINS observations employed ABBA nodding sequences along the slit to facilitate sky background subtraction. Spectral reductions were performed using the IGRINS Pipeline Package \citep[IGRINS PLP;][]{Lee:2017}, including flat-field correction, A-B frame subtraction, telluric correction, and wavelength calibration using sky OH emission lines \citep{Han:2012, Oh:2014}. Telluric lines were removed by dividing the science spectra by those of  fast-rotating late-B to early-A dwarfs observed near in time and air mass. The spectral orders were normalized, stitched together, and combined after excluding low S/N edges using {\tt iraf} \citep{IRAF}. Finally, the spectra were shifted to laboratory wavelengths in air, correcting for stellar radial velocities. Significant effort was devoted to defining local continua in spectral segments containing the lines of interest to address residual modulations in continuum levels. Typical spectra around the spectral lines used in this study are shown in Figs. \ref{fig:A1}-\ref{fig:A5}.

 \begin{figure*}
  \includegraphics[trim=0cm 0cm 0cm 0cm,clip,width=1\textwidth, angle=0]{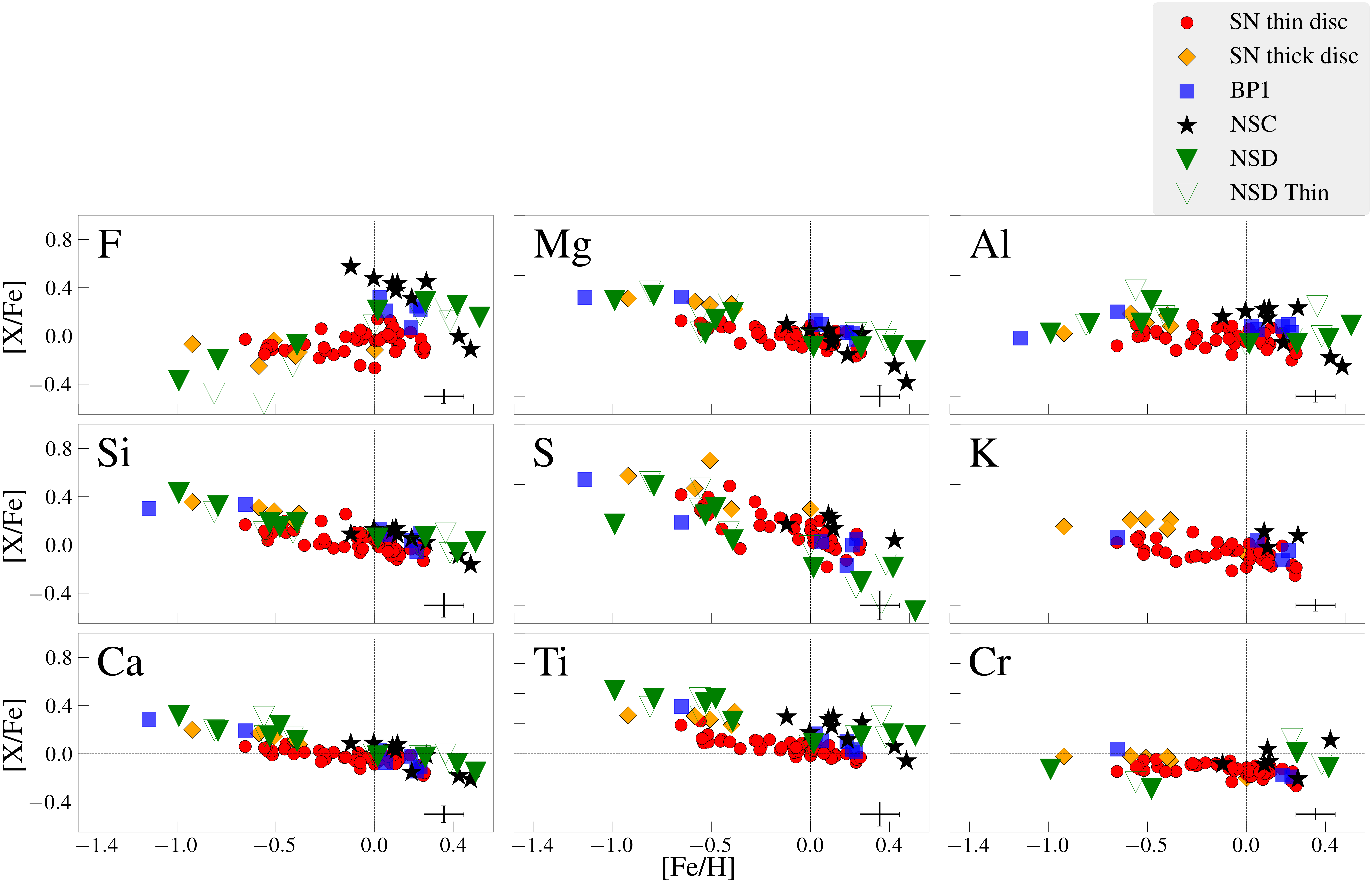}
  \caption{Abundance ratios versus metallicities for F, Mg, Al, Si, S,  Ca, Ti, and Cr for stars in the Nuclear Stellar Disc (green triangles) are shown. Comparison populations are also included: {\it(i)} solar-neighborhood thin-disc (denoted "SN thin disc" in the legend) and thick disc ("SN thick disc") stars are represented by red circles and yellow diamonds, respectively; {\it(ii)} stars observed in the inner-bulge field located at $1^\circ$ north of the Galactic Center -- $(l, b) = (0^\circ, +1^\circ)$ -- are shown as blue squares (denoted with "BP1" , i.e. BP1 = "B Plus 1"); and {\it(iii)} stars observed in the Nuclear Star Cluster are indicated by black star symbols. See text for references. The green open triangles (denoted "NSD Thin") represent the abundances for the NSD stars using alternative stellar parameters determined with the assumption that oxygen abundances follow the thin-disc trend (see text), indicating typical uncertainties. We present the abundance panel for potassium, K,  to demonstrate consistency with the figures for the NSC in \citet{NSC_all:25}. However, we were unable to obtain reliable K abundances for the NSD.} 
  \label{fig:alpha}%
\end{figure*}

 \begin{figure}
  \includegraphics[trim=0cm 0cm 0cm 0cm,clip,width=1\columnwidth]{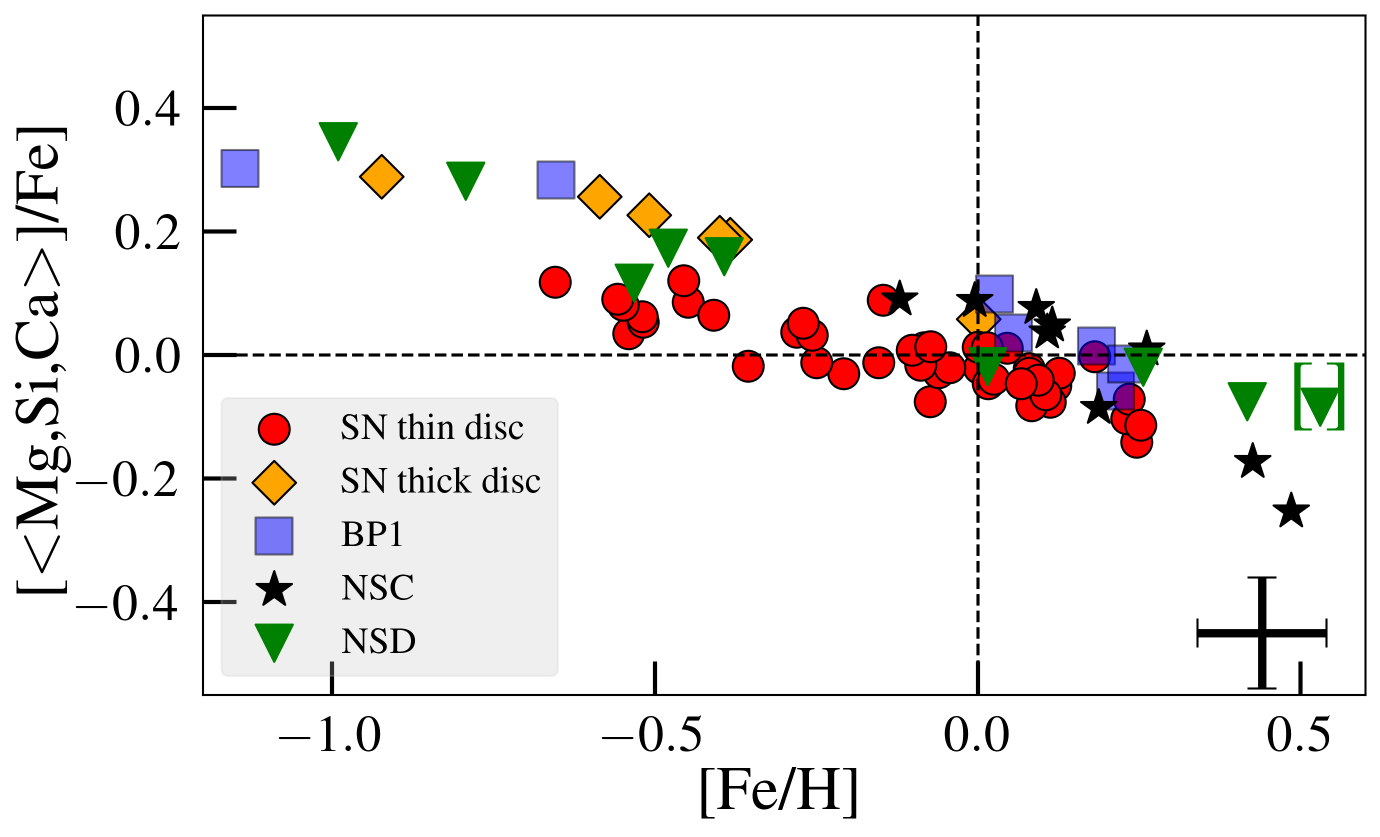}
  \caption{Simple mean of the three $\alpha$ elements Mg, Si, and Ca for the NSD (green triangles). Also shown are the  comparison samples of the Galactic disc populations (red circles and yellow diamonds), the inner-bulge  (blue squares), and the NSC (black star symbols). Typical uncertainties are given in the the lower right corner. The most uncertain determination of our abundance is for our coolest, most metal-rich, and faintest star. This is marked with a bracket.  
  }   
  \label{fig:alphamean}%
\end{figure}

 \begin{figure*}
  \includegraphics[trim=0cm 0cm 0cm 0cm,clip,width=1\textwidth]{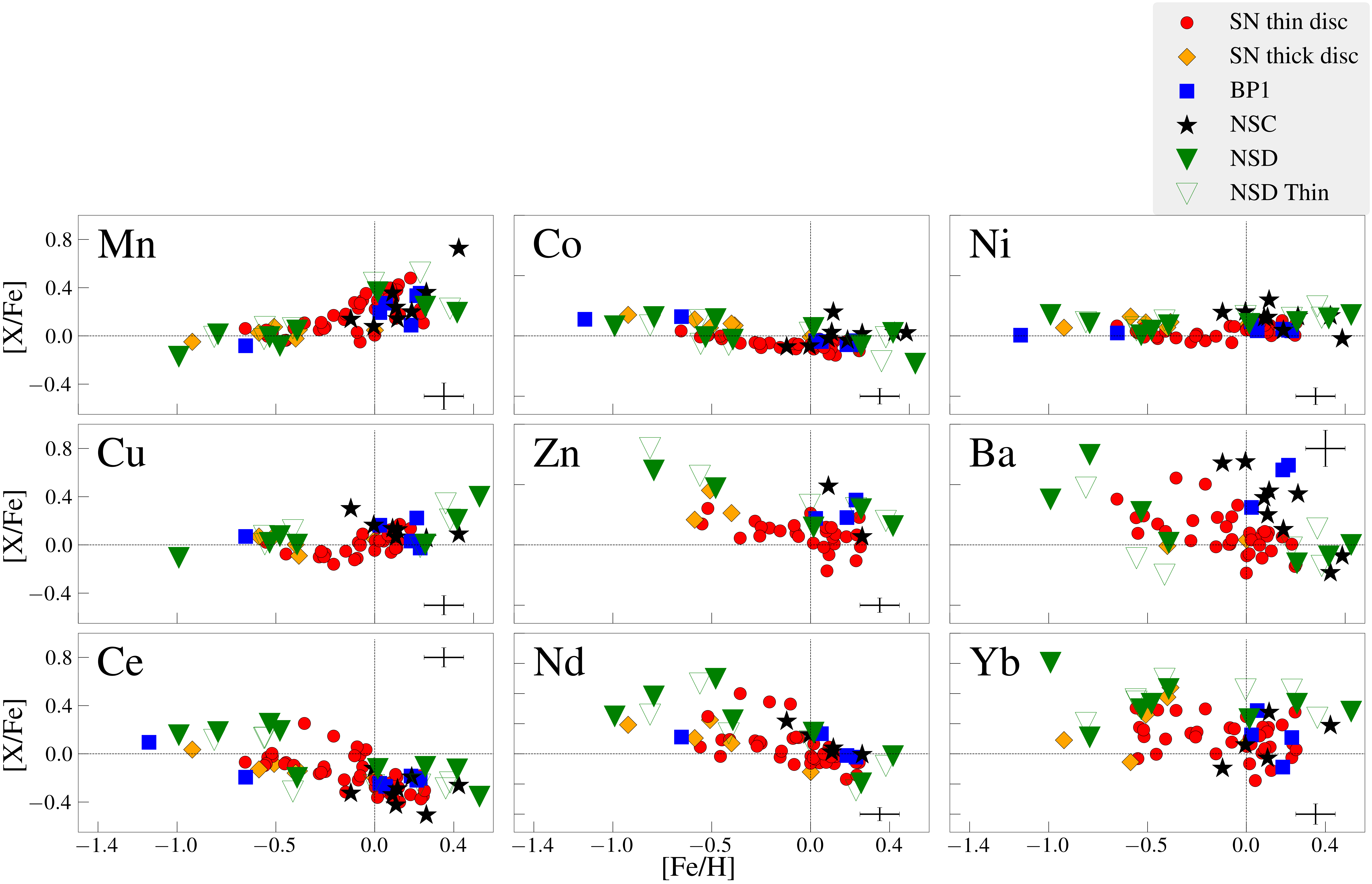}
  \caption{Abundance ratios versus metallicities for Mn, Co, Ni, Cu, Zn, Ba, Ce, Nd, and Yb for stars in the Nuclear Stellar Disc (green triangles) are shown. Thin-disc and thick-disc stars from the solar neighborhood are represented by red circles and yellow diamonds, respectively;  stars observed 1 degree north of the Galactic Center are shown as blue squares; and stars observed in the Nuclear Star Cluster are indicated by black star symbols. The green open triangles represent the abundance ratios for the NSD stars using 
  alternative stellar parameters determined under the assumption that the oxygen abundances follow thin-disc trend (see text), indicating typical uncertainties. See text for references.
 }    
  \label{fig:heavy}%
\end{figure*}

\section{Analysis}
\label{sec:analysis} 
 
We have derived the stellar parameters and abundances of a range of elements for  stars in the NSD using the same method as that applied in the studies of the populations we have used for comparison with the NSD trends, namely \citet{Nandakumar:2023,Nandakumar:24_21elements,nandakumar:24,ryde:25,NSC_all:25}.  
This differential comparison ensures that the systematic uncertainties are kept to a minimum. The derived abundances are given in the solar abundance scale of \citet{solar:sme}.

In \citet{Nandakumar:2023}, this iterative, spectroscopic method
was introduced for determining the stellar parameters\footnote{Effective temperature (\teff), surface gravity (\logg), metallicity (\feh), and microturbulence ($\xi_\mathrm{micro}$)} of M giants with \(3350 < T_{\rm eff} < 4000\,\mathrm{K}\) using high-resolution, H-band spectra and isochrones. In short, the method uses temperature-sensitive OH molecular lines to constrain the effective temperature, assuming a classification into thin- or thick-disc populations for an estimation of the oxygen abundances. A range of iron, CO, and CN lines are further used to determine the \feh, C, and N abundances as well as the microturbulence. The surface gravity is then found from isochrones for a given set of temperature and metallicity. A new iteration with the updated parameters is then run, and this process is repeated until convergence is achieved.  The stellar parameters of our stars are presented in Table \ref{table:parameters}, and range $3460<$ \teff$< 3840$\,K, $0.5<$ \logg$< 1.5$, $-1.0<$\feh$<+0.5$, and $1.5<\,$\microturb$<2.5$. The C and N abundances were determined to ensure that the CO and CN lines fit the spectra, as these often blend with the lines of interest for abundance determinations. Notably, for two of our metal-poor stars, a significant enhancement in N is required for these lines to match. 

The method for determining the stellar parameters was validated in \citet{Nandakumar:2023} by comparing its results with benchmark stars and literature values, showing good agreement and robustness against assumptions in initial parameters.
It was also successfully applied to open cluster stars by \citet{shilpa:24a}, demonstrating its accuracy in metallicity determination for open cluster star members, which are expected to have consistent metallicities regardless of other stellar parameters. Since the assumption
of the population or the oxygen abundance is the main source of uncertainty in the method, typical uncertainties were estimated by testing the impact of misclassifying stars into the wrong stellar population, thereby adopting an incorrect [O/Fe] value. Varying [O/Fe] by $\pm 0.2$\,dex resulted in uncertainties of $\pm$ 100 K in \teff, $\pm$0.2 dex in \logg, $\pm$0.1 dex in \feh, and $\pm$0.1\,\kms\ in $\xi_\mathrm{micro}$. A change in the assumed isochrone age (2 Gyr vs. 10 Gyr) altered 
\logg\ by only $0.1$\,dex. 

For the uncertainties in the derived abundances arising from the  stellar parameter uncertainties, we followed the method described in \citet{Nandakumar:2023}, \citet{Nandakumar:24_21elements}, and \citet{ryde:25}, where we redetermined the abundance from each selected line by generating 50 realizations of the stellar parameters. The stellar parameters were randomly drawn from normal distributions centered on the adopted stellar values, with typical uncertainties taken as the standard deviations. The uncertainty for each line was defined as half the difference between the 84th and 16th percentiles of the resulting abundance distribution. The final abundance uncertainty for each element was taken as the standard deviation of the mean of the individual line uncertainties. The mean uncertainties in the derived abundances range from 0.05 to 0.15.

Trends for 21 elements were determined from IGRINS spectra of M giants in the solar neighborhood in \citet{Nandakumar:24_21elements} and \citet{nandakumar:24} succeeded in determining the same number of elements in obscured stars 1 degree north of the Galactic Center. \citet{ryde:25} and \citet{NSC_all:25} were able to determine 19 elements in NSC stars.  Among these elements are the diagnostically important neutron-capture elements; Ce (s/r=85/15)\footnote{The solar ratio is from \citet{bisterzo:14}, but it can vary over time and with metallicity.}, and Nd (s/r=60/40) can be derived from H-band spectra, as first demonstrated in APOGEE spectra by \citet{Cunha:2017} and \citet{Hayes:2022}, respectively. Additional neutron-capture elements can be derived from  the higher-resolutions IGRINS spectra, 
as shown by \citet{montelius:22} in the case of Yb (s/r=40/60) and by \citet{Nandakumar:ba} in the case of Ba (s/r=90/10). Phosphorus (P) can also be determined, particularly in K giants, though it becomes increasingly challenging as the stars get cooler  \citep{nandakumar:22}. Here, we will use the recommended spectral lines from \citet{Nandakumar:24_21elements} to determine the same elements discussed for the NSC \citep{NSC_all:25}. We carefully define individual local continua around all spectral lines.

As in the analogous analyses mentioned above, we will synthesise our spectra using Spectroscopy Made Easy (SME) \citep{sme,sme_code}, which models the spherical radiative transfer through a stellar atmosphere model, selected by interpolating in a grid of one-dimensional (1D) Model Atmospheres in a Radiative and Convective Scheme (MARCS) stellar atmosphere models \citep{marcs:08}. The atomic and molecular line data are the same as described in \citet{Nandakumar:24_21elements}. Furthermore, we use the same  departure coefficients from non-local thermodynamic equilibrium (non-LTE) grids for the elements C, N, O, Na, Al, Mg, Si, Ca, Ti, Mn, Fe, Cu, and Ba \citep{amarsi:20,amarsi:22}.

\section{Results}
\label{sec:results}

\begin{figure*}
  \includegraphics[width=\textwidth]{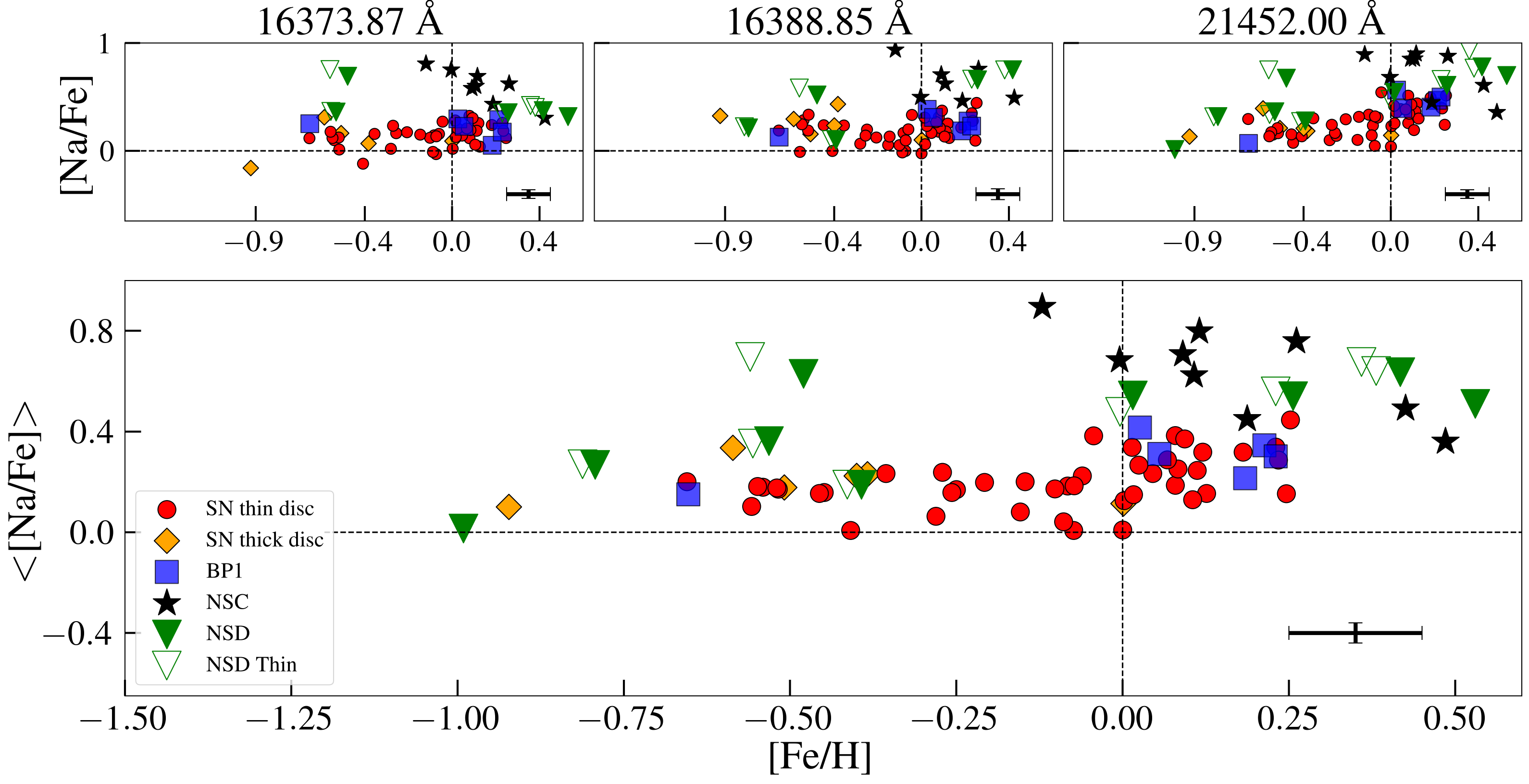}
  \caption{[Na/Fe] versus [Fe/H] for different stellar populations. The Nuclear Stellar Disc stars are represented by green open and filled triangles, NSC stars by black star symbols, inner bulge stars by blue squares, solar neighborhood thin-disc stars by red filled circles, and thick-disc stars by orange diamonds. In the upper panels, the trends from the individual spectral lines used in the analysis are shown, while the mean trend is displayed in the lower panel. Not all three lines were usable for every star. The green open triangles (denoted "NSD Thin") represent the abundances for the NSD stars using alternative stellar parameters determined with the assumption that oxygen abundances follow thin-disc trend (see text), indicating typical uncertainties.
}   
  \label{fig:na}%
\end{figure*}

We present our results in Table \ref{tab:element_abundances}  and in Figs. \ref{fig:alpha} -- \ref{fig:na}, showing abundance ratios for 18 elements as a function of metallicities for nine  stars in the Nuclear Stellar Disc, represented by green triangles. For comparison, we also include the abundances derived for solar neighborhood stars from the studies by \citet[][shown as red circles and yellow diamonds]{Nandakumar:2023,Nandakumar:2023b,Nandakumar:ba,Nandakumar:24_21elements}. Additionally, we present the abundances for stars in the inner bulge from the study by \citet[][represented as blue squares]{nandakumar:24}, as well as for stars in the Nuclear Star Cluster from the studies by \citet[][shown as black star symbols]{ryde:25,NSC_all:25}. 

In the figures, estimates of the uncertainties in the abundance 
are indicated as error bars in the lower right corners of each trend. These uncertainties for a typical star are also provided in the last column in Table \ref{tab:element_abundances}.  The true uncertainties are difficult to estimate since they depend on a range of sources, such as unknown blends, spurious spectroscopic features, such as DIBS, and residuals from the elimination of the telluric lines. The uncertainties due to the uncertainties in  the stellar parameters can, however, be assessed, and these are all of the order of 0.1\,dex (0.05-0.15), see Sect. \ref{sec:analysis}. A valid uncertainty estimate is the general dispersion of the trends, assuming that the cosmic trends are narrower than the uncertainties. 
As an additional indication of  the uncertainties in the abundance ratio determinations for our NSD stars, we also display the abundances derived using the stellar parameters determined with the assumption that oxygen abundances follow thin-disc trend \citep[green open triangles; for further details, see][]{Nandakumar:24_21elements,NSC_all:25}. We note that this variation in these abundances are within the uncertainties marked as error bars in each panel. In general, the more metal-rich and the cooler a star is, the more challenging spectral analysis becomes due to increasing molecular blending and the presence of unknown atomic blends. This should be considered when comparing abundances determined for metal-poor   and metal-rich stars, the latter of which are often cool. NSD\_25 is an example of such a metal-rich and cool star, with \teff=3460 and a [Fe/H]$=+0.51$, for which higher uncertainties can be expected, but difficult to assess. 

Among our nine  NSD stars, five have sub-solar metallicities, whose abundances can be compared to the two stars in the inner-bulge sample with sub-solar metallicities, as well as to the thin- and thick-disc samples of solar neighborhood stars. The remaining four NSD stars have solar or super-solar metallicities, extending up to \feh$=+0.5$, similar to the NSC stars. Up to \feh$=+0.25$, our NSD stars can be compared to all the comparison populations.

For the NSD stars, the $\alpha$-element trends (Mg, Si, S, and Ca) and that of Ti all closely follow the thick-disc solar neighborhood trends (yellow diamonds in Fig. \ref{fig:alpha}), as well as the inner-bulge star trend (blue squares).
This alignment is demonstrated clearly in the mean 
$\alpha$-trend shown in Fig. \ref{fig:alphamean}, where we present a simple mean of the three best determined $\alpha$ elements: Mg, Si, and Ca. The NSD trend follows the thick-disc pattern at subsolar metallicities, similar to the inner-bulge stars.\footnote{Denoted BP1 in the figures, indicating the field located $1^\circ$ north of the Galactic Center, $(l, b) = (0^\circ, +1^\circ)$, i.e. BP1 = "B Plus 1"} At super-solar metallicities, the NSD, as well as the inner-bulge and NSC stars, follow the upper envelope of the metal-rich solar neighborhood trend. The determination of sulfur abundances is affected by larger uncertainties, but sulfur is expected to behave as a normal  $\alpha$ element \citep[see, for example,][]{ryde:04,ryde:06,matrozis}.

At the highest metallicities ([Fe/H]>0.25), where we unfortunately lack solar neighborhood comparison stars, the two NSD stars show higher  $\alpha$ abundance values than the two NSC stars, see Fig. \ref{fig:alphamean}. Although this difference is within the uncertainties, it is systematic for all $\alpha$ elements. It should, however, be noted that the most metal-rich NSD star (NSD\_25) is cool, leading to higher expected uncertainties. We have greater confidence in the abundance determination of the two NSC \citep[from][]{ryde:25}, as they are warmer (\teff$=3709$ and $3825$\,K) and brighter (H=13.4 and 13.9) compared to NSD\_25 (\teff=$3460$\,K and H=15.0), which is the coolest and faintest star in our sample, see Table \ref{table:obs} and \ref{table:parameters}. The higher effective temperatures and H magnitude ensure a more accurate determination of the stellar parameters and, consequently, the abundances of the NSC stars.   More data is needed to examine this metallicity region in greater detail. 

The trend for the odd-Z element Al, shown in Fig. \ref{fig:alpha}, also closely follows the thick-disc and inner-bulge trends. Especially for the metal-poor stars, the Al trend decreases with decreasing metallicities, as expected from chemical evolution models \citep[see, e.g.,][]{Adibekyan:12,smiljanic:16,kobayashi:20}. At super-solar metallicities, the NSD trend does not significantly differ from the inner-bulge and NSC trends, remaining within their scatter. We were not able to measure any reliable K abundances for the NSD stars due to spurious features in the spectral region around the two K lines at $15\,163-15\,168$\,\AA, see Fig. A2. For consistency, we still present the K trends for the comparison populations in Fig. \ref{fig:alpha}.

The iron-peak elements Cr (Fig. \ref{fig:alpha}), as well as Co and Ni (Fig. \ref{fig:heavy}) display generally flat trends across all populations, with the subsolar NSD trends closely following the thick-disc and inner-bulge trends. Co shows the largest difference between thin- and thick-disc stars, with Ni displaying a similar but smaller difference \citep{lomaeva:19,Nandakumar:24_21elements}. Our inner-bulge, NSC, and NSD stars are all consistent with the enhanced thick-disc trend. The abundances derived for Mn (Fig. \ref{fig:heavy}) in the different populations agree within uncertainties. However, there is a large discrepancy for the only two stars at the highest metallicities. Unfortunately, we do not have any solar neighborhood stars at these metallicities to compare to. More data is needed to investigate the trends at these high metallicities further.

Zn, displayed in Fig. \ref{fig:heavy}, is difficult to measure accurately, but since it behaves similarly to an 
$\alpha$  element \citep[see, e.g.,][]{Nandakumar:24_21elements}, the NSD stars, along with the inner-bulge and NSC stars, appear to follow the upper envelope of the thin-disc solar neighborhood trend, consistent with the behavior of the other $\alpha$ elements.

Cu, also displayed in Fig. \ref{fig:heavy}, is synthesized through the weak s-process in massive stars \citep[e.g.][]{pignatari:10} and exhibits an N-like trend in the solar neighborhood, generally increasing with metallicity, particularly for the thick disc \citep[see, e.g.,][]{forsberg:phd}. The subsolar NSD stars align well with the thick-disc trend, and  at super-solar metallicities, Cu abundances are consistent with those in inner-bulge and NSC trends, extending the disc trend. 
 
We have also successfully measured abundances for neutron-capture elements, namely Ba, Ce, Nd, and Yb (Fig. \ref{fig:heavy}). These elements are produced through a combination of the s- and r-processes. For Ba and Ce, the s-process dominates, with s/r ratios of 90/10 and 85/15, respectively, in the solar isotopic mixture \citep{bisterzo:14,prantzos:20}. Nd has a 60/40 ratio, indicating that it is still mainly an s-process element. Yb has a ratio close to 50/50 \citep{kobayashi:20,prantzos:18}, making it the most r-process-rich element among those investigated here. In the solar neighborhood, the thick-disc  trends for these s-process elements generally show a decreasing pattern with metallicity, following the lower envelope of the thin-disc banana-like distribution \citep{forsberg:phd}. Ba and Nd as we measure them in the NSD stars generally exhibit this behavior, while Ce shows larger scatter at subsolar metallicities, though it remains consistent with this trend. The Yb abundances are particularly challenging to measure, requiring a high S/N ratio, as the line lies in the wing of a strong CO line \citep{montelius:22}. Nevertheless, the thick-disc trend is expected to cut through the thin disc, resembling a tilted $\alpha$-element trend. More stars are needed to draw firm conclusions about the behaviour of the neutron-capture elements.

The fluorine abundance trends were found to be higher in the NSC than in the inner bulge by \citet{NSC_all:25} for their investigated stars, all of which have super-solar metallicities. They concluded that the temperature sensitivity of the HF lines, from which the fluorine abundance was derived, made the determination uncertain.
Here, we observe that fluorine follows the thick-disc trend at subsolar metallicities and aligns more closely with the inner-bulge trend where comparisons are possible, see Fig. \ref{fig:alpha}. The fluorine abundances remain higher than those of the thin-disc trend. Two of our stars, NSD\_1 and NSD\_19, which have metallicities of [Fe/H]$\sim-0.5$, have such weak lines that the fluorine abundances are unmeasurable and would likely have yielded very low values (see Table \ref{tab:element_abundances}). If these low values are significant or not has to be investigated further, with more stars. We note that these two stars are also the stars that needed very high N abundances in order to fit the CN and CO lines in the stellar-parameter determinations, see Sect. \ref{sec:analysis}. The nucleosynthetic contributions to the cosmic fluorine budget at these metallicities remain under debate \citep[see, e.g.,][]{guerco:22b,Nandakumar:2023b}.
In order to make any firm conclusions on the fluorine evolution in the different populations, more stars with higher S/N ratios in the H-band are needed to increase the accuracy of the temperature determination. This would enable high accuracies in the determined fluorine abundances. 

The only element trend in the NSC shown by \citet{NSC_all:25} to be significantly different -- by a factor of two -- compared to the inner bulge and discs was Na. Here, we also find enhanced Na abundances for the NSD stars with super-solar metallicities (see Fig. \ref{fig:na}), indicating a similarity between the NSC and NSD that differs from the inner-bulge stars. At subsolar metallicities, there is some scatter, but the trend is not significantly different from the thick-disc trend.

\section{Discussion}
\label{sec:discussion}

The extreme dust obscuration at optical wavelengths along the line of sight toward the Galactic Center prohibits optical analyses, but near-IR radiation can still be detected. High-resolution H and K band spectra, which are essential for accurate and precise abundance determinations, can be observed with a good S/N ratio using, for instance, the IGRINS spectrometer on the 8-meter Gemini telescopes, as demonstrated by the spectra analyzed here. Our analysis focuses on bright M giants, which require special attention to the methodology.  An essential aspect of our method is minimizing systematic uncertainties when comparing different stellar populations. Therefore, our abundance trends are compared  differentially with  stars from well-characterized populations in the solar neighborhood \citep[data from][]{Nandakumar:2023,Nandakumar:24_21elements} as well as  with stars in the inner bulge \citep[$1^\circ$ north of the Galactic Center as found by][]{nandakumar:24}, and stars in the NSC \citep{ryde:25,NSC_all:25}.  These comparison stars were observed using the same observational setup and analyzed with the same methodology, including an identical stellar parameter scale. 
Thus, together with the 
works of  \citet{ryde:25,NSC_all:25}, it is clear that it is now possible to investigate abundance trends from high-resolution, near-IR spectra in all locations of the Milky Way, even the extremely dust-obscured regions. 

It should be noted that the sets of stars in the different populations we investigate are selected from  lists of the brightest M giant stars in their fields and are not representative of the underlying Metallicity-Distribution Function (MDF). More stars and a careful understanding of the selection functions are needed to study the similarities and differences of the MDFs of the different populations. Whether the fact that we do not detect any metal-poor stars in our NSC sample presented in \citet{ryde:25}, while approximately half of the stars in the NSD sample have subsolar metallicities, could be due to a selection effect where metal-poor stars in the NSC are too faint to be detected. We know for a fact that metal-poor stars exist in the NSC \citep[see, for example,][]{ryde:2016_metalpoor}.

More data than we have presented in this paper is obviously needed in order to make firmer conclusions of the stellar populations and the the structures they belong to, but what our data (except Na) of only nine  NSD stars already shows is striking similarities, within uncertainties, between the abundance-ratio trends found in the NSD with those found in the inner bulge \citep[$1^\circ$ north of the Galactic Center as found by][]{nandakumar:24} for all metallicities and with those of the local thick disc for subsolar metallicities.
This is also true for the trends of the NSD and those of the NSC \citep{ryde:25,NSC_all:25}, spanning mostly supersolar metallicities. 

There is, for example, no indication of very different $\alpha$-element trends (Mg, Si, S, and Ca), as would have been expected if the star-formation rate and evolution would have been significantly faster \citep[see, e.g.,][]{matteucci_brocato:90}. To what extent our findings can constrain the early star-formation histories in the NSD and NSC, only full Galactic Chemical Evolution models can reveal, as they integrate the complex interplay of gas inflows, star-formation rates and stellar yields \citep[see, e.g.,][]{matteucci:12,friske:23}. The star-formation histories seem to be different in the NSD and the NSC \citep{schodel:20,lara:20,lara:21}, but we do not detect any significant differences in the abundance trends within the precision of our observations.  

Furthermore, the similarity across the wide range of elements we find,  with diverse nucleosynthetic paths and evolutionary timescales strengthens the conclusion of the chemical similarities between the populations investigated.
Especially, elements that evolve on different timescales than the commonly used $\alpha$ elements (fast evolution) are providing new insights. Examples of such elements with a slower evolution are the slow neutron-capture elements (mainly from AGB stars, low- to intermediate-mass stars) and manganese (mainly from Type Ia supernovae, SNe Ia). These stars release their products with a time delay compared to Type II SNe. We do not detect any significant differences in the trends for these elements.

Sodium is the only element displaying a distinct trend, with enhanced abundances in the NSD and NSC compared to thin disc stars, but also to inner-bulge stars.  As discussed in \citet{NSC_all:25}, there is no reason to believe that there should be any significant systematic uncertainties in our Na abundance determinations. 
A straight forward explanation for this finding is still lacking, based on the understanding of different Na sources, such as those in \citet{kobayashi:20}. It is, however,  interesting to note that high Na abundances were also found in the super-solar population in Liller 1, a complex, multi-population stellar system in the Galactic bulge \citep{liller1:24}. Furthermore, \citet{munoz:2018} found a Na spread in NGC 6528, which is  a metal-rich globular cluster \citep[\feh$\sim +0.1$; ][]{carretta:01}. Similar to what we find, they did not find an Na-O anti-correlation, nor an Mg and/or Al spread. 
\citet{carretta:01} also found a large Na excesses, by more than a factor of 2, but normal oxygen abundances for stars in the same cluster, NGC 6528.
Whether a connection can be found between the evolutionary histories for the NSD and these objects, remains to be determined.


Since we detect NSD stars down to \feh$=-1$, our data demonstrates that one may start addressing questions such as whether the signatures from accreted globular clusters \citep[see, e.g.,][]{carretta:09,bastian_lardo_GC} could be detected in the NSD population. Within our very limited sample, the five metal-poor stars show no signs of the combination of high Na abundances  with very low oxygen abundances nor high Al and low Mg, which are expected in globular clusters.  We note that this is also true for the two of our stars that show high N and low C abundance combinations, which can also be found in globular clusters. 
Thus, we do not observe any stars with obvious anti-correlations signatures, characteristic of globular clusters. 
A dedicated study of a much large sample of stars, 
along with a tailored formation model, could provide firmer constraints on the potential contribution of the formation channel from accreted globular clusters in the Galactic Center populations. 

Similarly to what was discussed in \citet{nandakumar:24}, we note that  also the NSD abundance trends, similarly to the trends in the inner-bulge and NSC stars, exhibit trends that follow the inner-disc sequence \citep{Haywood:2013,Dimatteo:2016}, aligning with the high-[$\alpha$/Fe] envelope of the metal-rich, thin-disc population in the solar vicinity.
Considering the strong connection between the star-formation histories of the thick disc and the bulge \citep[see, e.g.,][]{haywood:18}, our results suggest that the early star-formation history -- reflected in the abundance patterns of the stars studied here -- is consistent across different spatial scales. This consistency spans from the several-kiloparsec scale of the thick disc to the few-kiloparsec scale of the bulge and extends to the innermost regions of the Milky Way, in the Galactic Center.

In order to chemically investigate the  inside-out star formation process in nuclear rings  suggested by \citet{sormani:22}, many more stars need to be observed, which might be possible with the upcoming surveys with the MOONS multi-objects spectrometer \citep{MOONS2020}. Furthermore, in order to chemically characterize the suggested recent star-formation and chemical evolution in the NSD \citep{lara:20}, young stars need to be observed, not only the old population that is probed by the abundances of the stars studied here.






\section{Conclusions}
\label{sec:conclusion}

In this paper, we present an unprecedented set of elements measured in stars lying in the Nuclear Stellar Disc (NSD).  By successfully detecting and analyzing nine M giants, we have demonstrated that it is now possible to obtain accurate and precise abundance ratio trends for stars there. 
We present trends for 18 elements ranging different nucleosynthetic origins and timescales, providing important insights into the discussion on and modeling of the origin and formation history of the NSD. 
With more data in the future -- to increase precision and statistics to also cover substructure -- fundamental questions about the formation history of the NSD can be addressed, providing a new dimension in the discussions.

We find 
similarities between a majority of the abundance-ratio trends with those found in the inner bulge \citep[$1^\circ$ north of the Galactic Center as found by][]{nandakumar:24} and those of the Nuclear Stars Cluster \citep{ryde:25,NSC_all:25}.  The trends for 
$\alpha$-elements, as well as Al, Cr, Mn, Co, Ni, Cu, Zn, and neutron-capture elements align well with the thick-disc behavior at subsolar metallicities. At super-solar metallicities, where we also can compare with the NSC, most elements follow the NSC and inner-bulge trends. The NSD population is thus consistent with the chemical sequence observed in the inner Galaxy (the inner-disc sequence) implying that the early star-formation history is similar across scales \citep{haywood:18}. We do not detect any stars exhibiting clear Na-O or Mg-Al anti-correlation signatures in our very limited data, which could have indicated the presence of accreted globular cluster stars at low metallicities. 

Sodium is the only element displaying a distinct trend, with enhanced abundances in the NSD and NSC at supersolar metallicities compared to thin disc stars, but also to inner-bulge stars.  High Na abundances have earlier been found in the metal-rich population in the complex, multi-population stellar system  Liller 1 in the Galactic bulge \citep{liller1:24} and a Na spread is found in the metal-rich globular cluster, NGC 6528 \citep{carretta:01,munoz:2018}, showing no anti-correlations.
Further investigations are needed to determine whether there are any evolutionary links between the NSD or NSC and the complex stellar system Liller 1 and metal-rich globular clusters.

Our abundance trends provide modelers of the Galactic Center with a new dimension of data. Our abundance trends can now contribute to discussions on the formation channels of the Galactic Center structures. Although our sample includes only nine stars, due to the challenging nature of the observations, this study, together with the preceding papers on the NSC \citep{ryde:25,NSC_all:25}, demonstrates that it is now indeed possible to obtain high-quality abundance data using high-resolution spectrometers on large telescopes. Our observations open new opportunities for abundance determinations in highly obscured regions of the Milky Way, setting a precedent and highlighting future possibilities. The upcoming MOONS survey \citep{MOONS2020}, although operating at a lower spectral resolution, will offer the potential to expand these studies further.

\begin{acknowledgements}
N.R.\ acknowledges support from the Swedish Research Council (grant 2023-04744) and the Royal Physiographic Society in Lund through the Stiftelsen Walter Gyllenbergs and Märta och Erik Holmbergs donations. G.N. acknowledges the support from the Crafoord Foundation via the Royal Swedish Academy of Sciences (Vetenskapsakademiens stiftelser; CR 2024-0034). R.A. acknowledges support from the Lise Meitner grant from the Max Planck Society (grant PI: M. Bergemann) and from the European Research Council (ERC) under the European Union’s Horizon 2020 research and innovation programme (Grant agreement No. 949173, PI: M. Bergemann). A.~R.~A. acknowledges support from DICYT through grant 062319RA and from ANID through FONDECYT Regular grant No. 1230731. M.Z. acknowledges support from the National Agency for Research and Development (ANID) BASAL Center for Astrophysics and Associated Technologies (CATA) through grant
and FB210003, and by the ANID Millenium Science Initiative, AIM23-0001, 
awarded to the Millennium Institute of Astrophysics (MAS). This work used The Immersion Grating Infrared Spectrometer (IGRINS) was developed under a collaboration between the University of Texas at Austin and the Korea Astronomy and Space Science Institute (KASI) with the financial support of the US National Science Foundation under grants AST-1229522, AST-1702267 and AST-1908892, McDonald Observatory of the University of Texas at Austin, the Korean GMT Project of KASI, the Mt. Cuba Astronomical Foundation and Gemini Observatory.
This work is based on observations obtained at the international Gemini Observatory, a program of NSF’s NOIRLab, which is managed by the Association of Universities for Research in Astronomy (AURA) under a cooperative agreement with the National Science Foundation on behalf of the Gemini Observatory partnership: the National Science Foundation (United States), National Research Council (Canada), Agencia Nacional de Investigaci\'{o}n y Desarrollo (Chile), Ministerio de Ciencia, Tecnolog\'{i}a e Innovaci\'{o}n (Argentina), Minist\'{e}rio da Ci\^{e}ncia, Tecnologia, Inova\c{c}\~{o}es e Comunica\c{c}\~{o}es (Brazil), and Korea Astronomy and Space Science Institute (Republic of Korea).
The following software and programming languages made this
research possible: TOPCAT (version 4.6; \citealt{topcat}); Python (version 3.8) and its packages ASTROPY (version 5.0; \citealt{astropy}), SCIPY \citep{scipy}, MATPLOTLIB \citep{matplotlib} and NUMPY \citep{numpy}.
\end{acknowledgements}

%
%


\bibliographystyle{aa}



\appendix
\begin{appendix} 

\section{Additional figures}

\begin{figure*}
  \includegraphics[width=\textwidth]{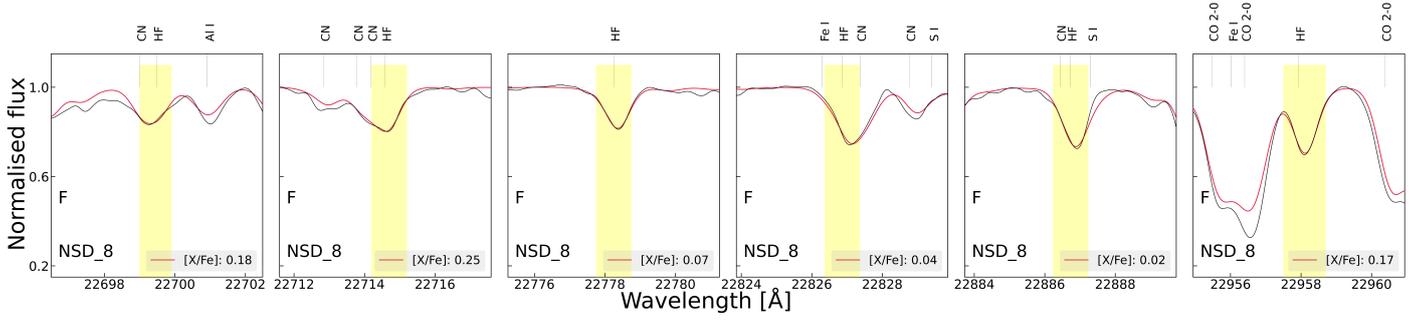}
  \caption{Sample spectra of the NSD star (NSD$\_8$) with wavelength regions centered at the individual absorption lines of fluorine. The black line in each panel denotes the observed spectrum, the crimson line denotes the best-fit synthetic spectrum, and the yellow bands denote the line masks defined for each absorption line. The abundance corresponding to the best-fit case
for each line are listed in every panel along with all identified atomic and molecular lines at the top of each panel.}   
  \label{fig:A1}%
\end{figure*}

\begin{figure*}
  \includegraphics[width=\textwidth]{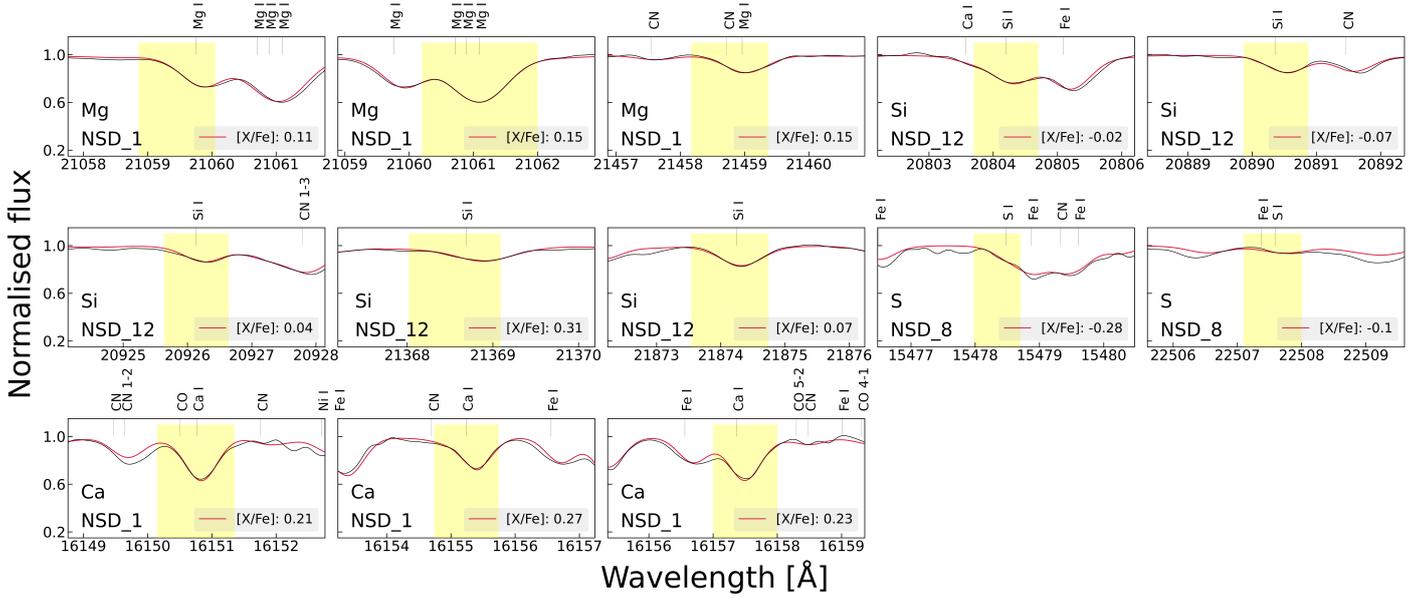}
  \caption{Sample spectra of the NSD stars (NSD$\_1$, NSD$\_8$ and NSD$\_12$) with wavelength regions centered at the individual absorption lines of the alpha elements- Mg, Si, and Ca. The black line in each panel denotes the observed spectrum, the crimson line denotes the best-fit synthetic spectrum, and the yellow bands denote the line masks defined for each absorption line. The abundance corresponding to the best-fit case
for each line are listed in every panel along with all identified atomic and molecular lines at the top of each panel.}   
  \label{fig:A2}%
\end{figure*}

\begin{figure*}
  \includegraphics[width=\textwidth]{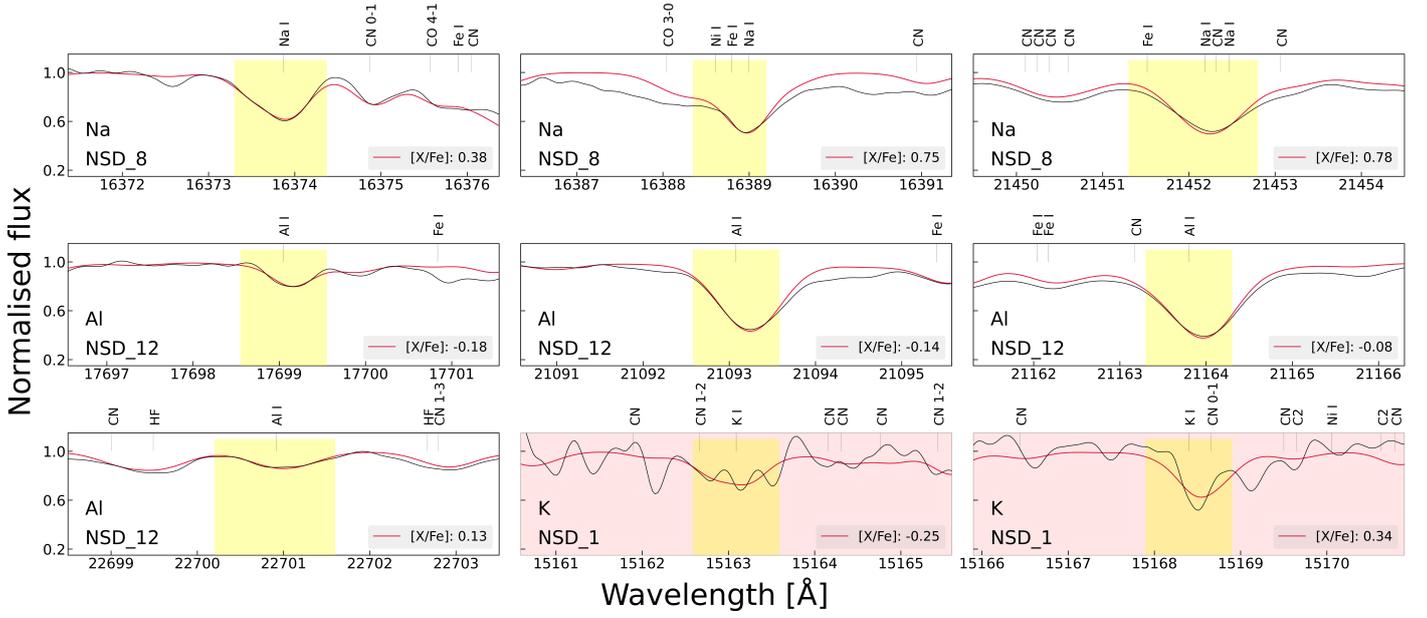}
  \caption{Sample spectra of the NSD stars (NSD$\_1$, NSD$\_8$ and NSD$\_12$) with wavelength regions centered at the individual absorption lines of the odd-Z elements- Na, Al, and K. Arrangement of figures and plot descriptions are similar to Fig. \ref{fig:A1}. Panels with red background indicate the lines that have not been used to determine abundances due to various factors such as high noise, bad telluric correction etc.}   
  \label{fig:A3}%
\end{figure*}

\begin{figure*}
  \includegraphics[width=\textwidth]{Ironpeak_spectra_NSD.pdf}
  \caption{Sample spectra of the NSD stars (NSD$\_1$, NSD$\_8$ and NSD$\_12$) with wavelength regions centered at the individual absorption lines of the iron-peak elements- Ti, Cr, Mn, Co and K. Arrangement of figures and plot descriptions are similar to Ni \ref{fig:A1}.}   
  \label{fig:A4}%
\end{figure*}

\begin{figure*}
  \includegraphics[width=\textwidth]{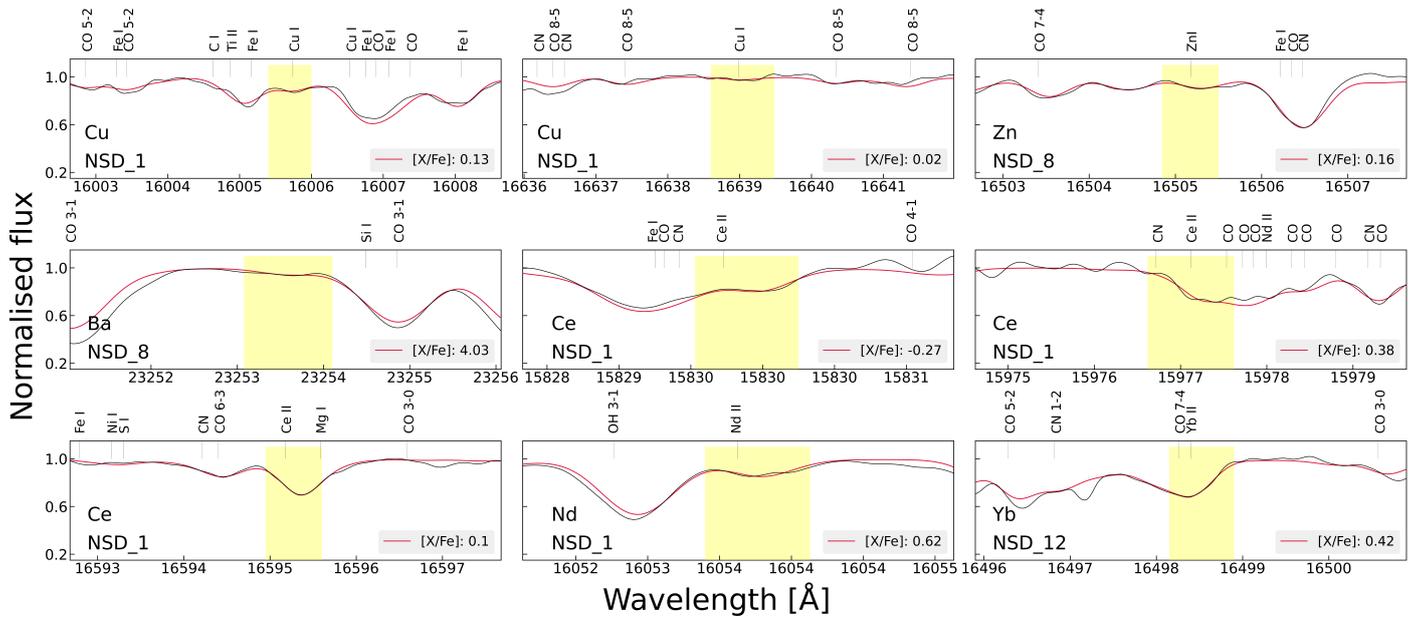}
  \caption{Sample spectra of the NSD stars (NSD$\_1$, NSD$\_8$ and NSD$\_12$) with wavelength regions centered at the individual absorption lines of the neutron-capture elements- Cu, Zn, Ba, Ce and Nd. Arrangement of figures and plot descriptions are similar to Fig. \ref{fig:A1}.}   
  \label{fig:A5}%
\end{figure*}

\end{appendix}

\end{document}